\documentclass[prl,twocolumn,letterpaper,superscriptaddress,floatfix,longbibliography]{revtex4-1}
\usepackage[space]{grffile}
\usepackage{bm,graphicx,amsmath,amssymb,color}
\usepackage{euscript,multirow,tabularx}
\usepackage{longtable}
\usepackage{float}
\usepackage[draft]{todonotes}
\usepackage{fancyvrb}
\usepackage{changes}

\usepackage{physics}
\usepackage{calrsfs}
\usepackage{footmisc}

\begin{document}

\title{Bright and dark states of two distant macrospins strongly coupled by phonons}

\author{K. An}
\email{Current address: Quantum Spin Team, Korea Research Institute of Standards and Science, Daejeon, Republic of Korea; kyongmo.an@kriss.re.kr}
\affiliation{Univ. Grenoble Alpes, CEA, CNRS, Grenoble INP, Spintec, 38054 Grenoble, France}

\author{R. Kohno}
\affiliation{Univ. Grenoble Alpes, CEA, CNRS, Grenoble INP, Spintec, 38054 Grenoble, France}

\author{A.N. Litvinenko}
\affiliation{Univ. Grenoble Alpes, CEA, CNRS, Grenoble INP, Spintec, 38054 Grenoble, France}

\author{R. L. Seeger}
\affiliation{Univ. Grenoble Alpes, CEA, CNRS, Grenoble INP, Spintec, 38054 Grenoble, France}

\author{V. V. Naletov} 
\affiliation{Univ. Grenoble Alpes, CEA, CNRS, Grenoble INP, Spintec, 38054 Grenoble, France}
\affiliation{Institute of Physics, Kazan Federal University, Kazan 420008, Russian Federation}

\author{L. Vila}
\affiliation{Univ. Grenoble Alpes, CEA, CNRS, Grenoble INP, Spintec, 38054 Grenoble, France}

\author{G. de Loubens}  
\affiliation{SPEC, CEA, CNRS, Universit\'e Paris-Saclay, 91191 Gif-sur-Yvette, France}

\author{J. Ben Youssef} 
\affiliation{Lab-STICC, CNRS, Universit\'e de Bretagne Occidentale, 29238 Brest, France}

\author{N. Vukadinovic}
\affiliation{Dassault Aviation, Saint-Cloud 92552, France} 

\author{G.E.W. Bauer} 
\affiliation{WPI-AIMR, IMR, and CSRN, Tohoku University, Sendai 980-8577, Japan}
\affiliation{Zernike Institute for Advanced Materials, University of Groningen, 9747 AG Groningen, Netherlands}

\author{A.~N. Slavin} \affiliation{Department of Physics, Oakland University, Michigan 48309, USA}

\author{V.~S. Tiberkevich} \affiliation{Department of Physics, Oakland University, Michigan 48309, USA}

\author{O. Klein}
\email[Corresponding author: ]{oklein@cea.fr}
\affiliation{Univ. Grenoble Alpes, CEA, CNRS, Grenoble INP, Spintec, 38054 Grenoble, France}

\def \C {{\mathcal C}}

\date{\today}

\begin{abstract}
 We study the collective dynamics of two distant magnets coherently coupled by acoustic phonons that are transmitted through a non-magnetic spacer. By tuning the ferromagnetic resonances of the two magnets to an acoustic resonance of the intermediate, we control a coherent three level system. We show that the parity of the phonon mode governs the indirect coupling between the magnets: the resonances with odd / even phonon modes correspond to out-of-phase / in-phase lattice displacements at the interfaces, leading to bright / dark states in response to uniform microwave magnetic fields, respectively. The experimental sample is a tri-layer garnet consisting of two thin magnetic films epitaxially grown on both sides of a half-millimeter thick non-magnetic single crystal. In spite of the relatively weak magneto-elastic interaction, the long lifetimes of the magnon and phonon modes are the key to unveil strong coupling over a macroscopic distance, establishing the value of garnets as a platform to study multi-partite hybridization processes at microwave frequencies.
\end{abstract}

\maketitle

\section{I. Introduction}

The coherent transfer of information between different wave forms is an important ingredient of quantum information processing as it allows interfacing, storage, and transport of quantum states \cite{wallquist2009hybrid,kurizki2015quantum,gu2017microwave,li2020hybrid}. The coupling between two distant quantum states, (1) and (2) can be coherently mediated by an intermediate third level (i), which may establish non-local entanglement \cite{bienfait2019phonon}. Such a tripartite hybridization, generates two collective modes under a uniform excitation field that are labeled bright or dark \cite{lukin2003colloquium,xu2008coherent,dong2012optomechanical}. Here we demonstrate control of the collective dynamics of two magnets by the polarity of the mutual coupling mediated by phonons through a high quality single crystal.

Strong coupling is possible only when the coupling rate between wave forms is larger than the geometric average of their relaxation rates \cite{reiserer2015cavity,spethmann2016cavity}, which corresponds to cooperativity parameters $\C>1$. Strong coupling between a microwave cavity mode and two superconducting qubits has long been realized \cite{wallraff2004strong,sillanpaa2007coherent}. The discovery of also strong coupling between a microwave circuit and ferrimagnetic objects \cite{soykal2010strong,huebl2013high,tabuchi2014hybridizing,zhang2014strongly,golovchanskiy2021approaching} has encouraged the study of coherent multilevel systems of two or more magnetic objects indirectly coupled by cavity photons, by both theory  \cite{yu2019prediction,grigoryan2019cavity} and experiments \cite{lambert2016cavity,bai2017cavity,xu2019cavity}. Studies focus on the dispersive regime, in which the two systems of interest are detuned and the coupling is mediated by virtual excitations of the intermediate level \cite{blais2007quantum,fragner2008resolving,filipp2011multimode,zhang2016cavity}. Moreover, until today, relatively little effort has been put to attempt to control the polarity of the coupling, which can permute the binding and anti-binding states.

Acoustic phonons in materials with low ultrasonic attenuation are excellent candidates to establish strong coupling with magnets as well \cite{bozhko2020magnon,li2021advances}. Acoustic phonon pumping via the magneto-elastic and -rotational couplings can lead to increased damping of the magnetization dynamics 
\cite{streib2018damping,sato2021dynamic,rezende2021theory}. Unidirectional phononic currents generated by magnetization dynamics have been predicted \cite{zhang2020unidirectional} and nonreciprocal propagation of acoustic phonons due to magneto-elastic coupling has been observed \cite{xu2020nonreciprocal,shah2020giant}. Coherent acoustic phonons can be also exploited to transfer incoherent spins \cite{ruckriegel2020long}. Magneto-elastic effects have been reported in various systems, such as metallic magnetic layers \cite{casals2020generation,peria2021magnetoelastic,zhao2020phonon,godejohann2020magnon}, nano-sized ferromagnets \cite{berk2019strongly}, dilute magnetic semiconductors \cite{kraimia2020time}, and layered structures \cite{latcham2020hybrid} as well as magnetic insulators \cite{an2020coherent,brataas2020spin,bozhko2017bottleneck}. In applications, the magneto-elastic coupling has also been used to design a low phase-noise magneto-acoustic resonator  consisting of a thin yttrium iron garnet (YIG) film epitaxially grown on gallium-gadolinium garnet (GGG) \cite{litvinenko2021tunable} or high-speed oscillating magnetic field sensors \cite{colombano2020ferromagnetic}. We reported coherent long-range transfer of angular momentum in YIG\textbar GGG\textbar YIG structures via circularly polarized acoustic waves propagating through a mm thick GGG spacer \cite{an2020coherent}. However, the magnetic resonances were detuned by more than the magneto-elastic coupling strength, thereby falling short of a full hybridization of the distant magnets.

In the following, we demonstrate hybridization of two macrospins, which establishes our system as a promising solid state platform for quantum information processing, allowing coherent communication between distant magnon states via phonons in a monolithic device.

We achieve full hybridization by careful tuning of both Kittel frequencies into a joint resonance with a phonon mode. At the degeneracy point, the magnon-phonon coupling induces a bonding-antibonding splitting that can be larger than the combined broadening, leading to bright and dark states of maximal and minimal coupling to an external readout antenna. We control the hybridization by either tilting the sample relative to an applied magnetic field, $H$, or by applying a vertical temperature gradient. Being able to select the index of the intermediate phonon mode offers an additional knob to control the parity of the coupling. We also detect the emerging bright and dark states electrically by Pt contacts on the YIG layers by spin pumping and the inverse spin Hall effect. The bright magnetic states emerge by the coupling to odd phonon modes. We observe a more than factor of 2 enhanced spin pumping signal, which is caused by the constructive interference of the amplitudes of the distant magnetic layers. On the other hand, by tuning the system to a dark state via an even index phonon mode, the microwave absorption and therefore the spin pumping signal are strongly suppressed.

\begin{figure*}
  \includegraphics[width=1\textwidth]{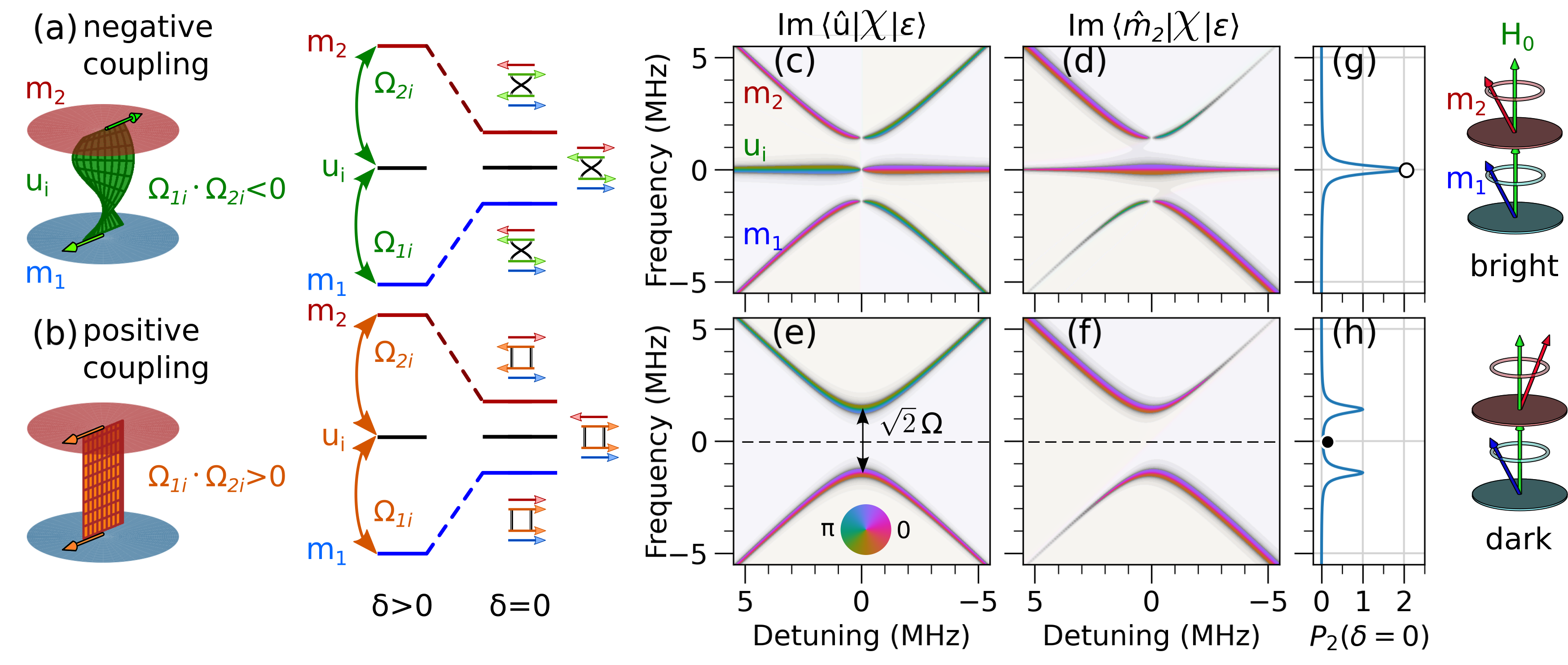}
  \caption{(Color online) Bright and dark hybridized magnon-phonon states or ``magnon polarons" in a YIG\textbar GGG\textbar YIG. Left: Out-of-phase negative (top row) /  in-phase positive  (bottom row) mutual coupling between two macrospins $m_1$ and $m_2$ in the outer YIG films when hybridized with an (a) odd  / (b) even index $i$ standing phonon mode $u_i$. $\delta=(\omega_2 - \omega_1)/2$  is half the splitting of the resonance frequencies of $m_1$ and $m_2$. Center: Amplitudes $\ket{\Psi} = (m_1,u_i,m_2)$ of the tripartite system as a function of the detuning \(\delta\), keeping  $(\omega_1 + \omega_2)/2 = \omega_i$ constant. The density plots (c)-(f) show in bi-variate colors (luminance:amplitude and hue:phase, cf. color wheel in (e)) the absorbed power in response to an excitation $\left\vert \hat \varepsilon \right\rangle=(1,0,1)/\sqrt{2}$ by (c and e) the acoustic subsystem $P_a \propto \mathrm{Im} \bra{\hat u }\chi\left\vert \varepsilon \right\rangle$ and  (d and f) the magnetic  top layer $P_{2} \propto \mathrm{Im} \bra{\hat{m_2}}\chi\ket{\varepsilon}$, with $\bra{\hat u} = (0,1,0)$ and $\bra{\hat{m_2}} = (0,0,1)$.  Right: (g) and (h) are the corresponding spectra of $P_2$ at $\delta=0$. The white/black circles indicate the bright/dark states observed at the triple crossing, see Eq.(\ref{eq:contrast}). The hybridized bright and dark magnetic states are illustrated on the right side of (g) and (h), respectively.}
  \label{fig_model}
\end{figure*}

\section{II. Model}

\begin{figure*}
  \includegraphics[width=1\textwidth]{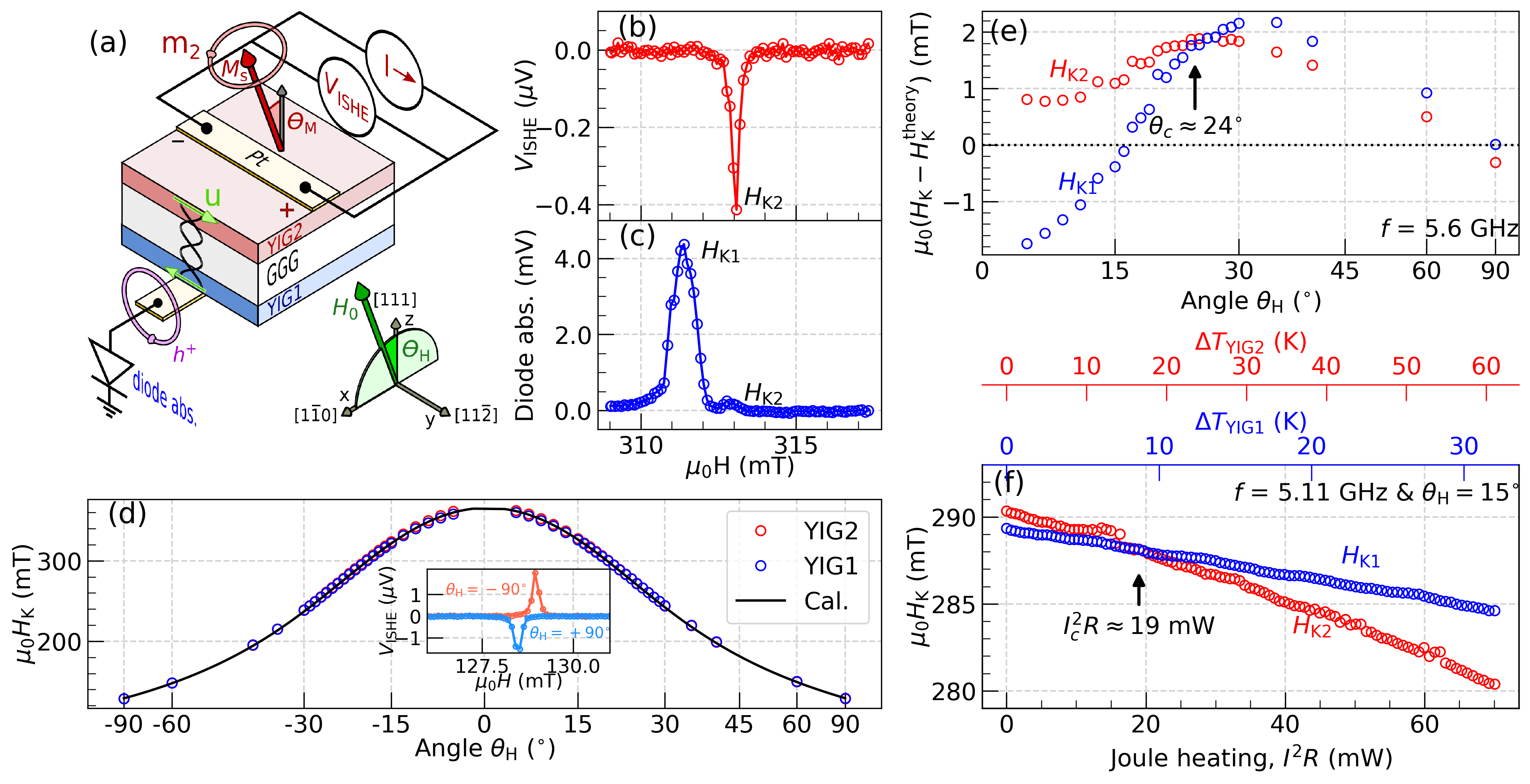}
  \caption{(Color online) Spin pumping experiment on a YIG1\textbar GGG\textbar YIG2 stack. (a) Schematic setup: A 300 $\mu$m-wide stripline antenna at the bottom couples inductively to the magnetic layers. A Pt strip deposited on top of the stack monitors the dynamics of the top (YIG2) layer. A DC current \textit{I} in the same Pt wire generates a temperature gradient by Ohmic heating. (b) Spin pumping voltage across the Pt ($V_\text{ISHE}$) and (c) microwave diode absorption signal. The peaks of (b) and (c) represent the Kittel resonances of YIG2 ($H_{K2}$) and YIG1 ($H_{\rm K1}$), respectively. Both spectra are taken at a polar angle $\theta_{H} = +17^{\circ}$. (d) $\theta_{H}$ dependence of the resonance fields of both layers. The solid line is a fit of $H_{\rm K2}(\theta_{H})$ by the Kittel formula, $H_K^{\rm theory}$  (see Appendix). The $V_{\rm ISHE}$ signal changes sign when rotating the magnetic field by 180$^{\circ}$, as shown in the inset \cite{fn2}. (e) Difference between observed and calculated resonance fields as a function of $\theta_{H}$. At the angle $\theta_{\rm c} \approx 24^{\circ}$ the resonances cross. (f) Variation of $H_{\rm K1}$ and $H_{\rm K2}$ as a function of Joule heating power in the Pt strip. The arrow indicates the Joule heating at which the detuning vanishes. The axes on top show the temperature rise of YIG1 and YIG2 deduced from the magnetizations as derived from the reduced resonance fields. Microwave power settings are as follows: 6 dBm for (b-e), and 0 dBm for (f), which is at the antenna resonance \cite{an2020coherent}.}
  \label{fig_setup}
\end{figure*}

We consider two macrospins with oscillating magnetic components $m_1$ and $m_2$ with eigenfrequencies $\omega_1$ and $\omega_2$, respectively. They are coupled by a standing acoustic wave with amplitude $u_i$ and frequency $\omega_{i}/(2\pi) =v/ \lambda_{i}$, where the integer $i$ is the mode index and $v$ is the sound velocity. $\lambda_{i}/2 = (2d+s)/i$ is half of the phonon wavelength that fits into the total crystal thickness, with \(d\) and \(s\) being the film thicknesses of the YIG and GGG films. The tripartite hybridization of the dimensionless state vector  $\ket{\Psi} =(m_1,u,m_2)$ is governed by the Hamiltonian (omitting the weak dipolar coupling): 

\begin{equation}
    \mathcal{H}_0=\begin{pmatrix}
    \omega_1 & \Omega_1/2 &  0 \\
    \Omega_1/2 & \omega_i  & \Omega_2/2  \\
    0 & \Omega_2 /2 & \omega_2 \\
    \end{pmatrix}.
\label{eq:coupling}
\end{equation}.

The phase shift of 0$^{\circ}$/180$^{\circ}$ at the free boundary conditions at the outer surfaces for phonon modes with even/odd index $i$ alternates the sign of the mutual coupling between  the top and bottom magnetic layers. We account it in Eq.\,(\ref{eq:coupling}) by $\Omega_{1} = -\Omega$ and $\Omega_{2} = (-1)^i \Omega_1$ \cite{an2020coherent}. 

At the degeneracy point ($\omega_1=\omega_i=\omega_2$) the eigenfrequencies relative to $\omega_i$ and corresponding normalized eigenvectors read:

\begin{equation}
    -\frac{\Omega}{\sqrt{2}}:
    \begin{pmatrix}
     1/2 & \\  1/\sqrt{2} \\ \pm 1/2 \\
    \end{pmatrix},\,
    0:
    \begin{pmatrix}
    1 /\sqrt{2} & \\ 0 \\ \mp 1/\sqrt{2}  \\
    \end{pmatrix},\,
    \frac{\Omega}{\sqrt{2}}:
    \begin{pmatrix}
      1/2 & \\  -1/\sqrt{2} \\ \pm 1/2 \\
    \end{pmatrix},
\label{eq:eigen}
\end{equation} respectively for even and odd indices (upper and lower sign).

In a bracket notation of the state vectors, the linear response to an excitation $\left\vert \varepsilon \right\rangle e^{j \omega t}$ oscillating at frequency $\omega/(2\pi)$  
\begin{equation}
   \left\vert \Psi \right\rangle =\chi \left\vert \varepsilon \right\rangle ,
\end{equation}
defines the dynamic susceptibility:
\begin{equation}
    \chi= \gamma (\omega {I}-\mathcal{H}_0)^{-1},
\end{equation}
where $I$ is the identity matrix and $\left\vert \varepsilon \right\rangle = \mu_0 h \,(1,0,1) / \sqrt{2}$ represents an external microwave magnetic field $h$ coupled inductively to the two macrospins, with $\mu_0$ being the vacuum permeability and $\gamma$ the gyromagnetic ratio.

Imaginary terms $j \eta _{s/a}$ \((j=\sqrt{-1})\) added to the diagonal elements of the Hamiltonian  broaden the energy levels, representing the magnetic and acoustic relaxation rates respectively. Fig.~\ref{fig_model}(d,f) and Fig.~\ref{fig_model}(c,e) show with a bi-variate color code the absorbed power under stationary conditions in the second layer and acoustic system under the microwave excitation, $P_2 = \kappa \, \mathrm{Im}\left\langle m_2\right\vert \chi\left\vert \varepsilon \right\rangle$ and $P_a =\kappa \,\mathrm{Im}\left\langle u_i\right\vert \chi \left\vert \varepsilon \right\rangle $, respectively, with $\kappa = \mu_0 M^2 \omega$. In these plots, the spectral integral of $P$ at fixed detuning are a definite-positive conserved quantity.  A negative \(P_2\) \cite{an2020coherent} (here in green) implies that the top layer returns power to the electromagnetic field when driven by an acoustic mode with phase shifted by 90$^\circ$ in advance of the microwave drive. The ``negative-power" acoustic modes change from even to odd (or vice versa) when crossing an acoustic resonance, because magnetic oscillations acquire a phase shift of ~180$^\circ$ that is added to the phase of the acoustic drive. The total power absorbed by the magnetic system vanishes for the dark state at the triple resonance, but does not become negative.

In Fig. 1(c-f), we plot the result  as a function of detuning $\delta \equiv (\omega_2 -\omega_1)/2$ between the magnetic layers  under the constraint $\omega_i=(\omega_1+\omega_2)/2$. When an even phonon mediates the interaction, we find an anticrossing. We point out, that the amplitude of the gap created by indirect coupling is $\sqrt{2} \, \Omega$, which is enhanced by $\sqrt{2}$ from the value produced by direct coupling. On the other hand, coupling by the odd phonon mode leads an additional central mode that does not depend on \(\delta\). The maximum value of $m_2$ is enhanced (compare the line cuts in Fig.~\ref{fig_model}(g) and (h)) by a constructive interference of the direct and indirect excitation amplitudes. 

\begin{figure}
  \includegraphics[width=0.5\textwidth]{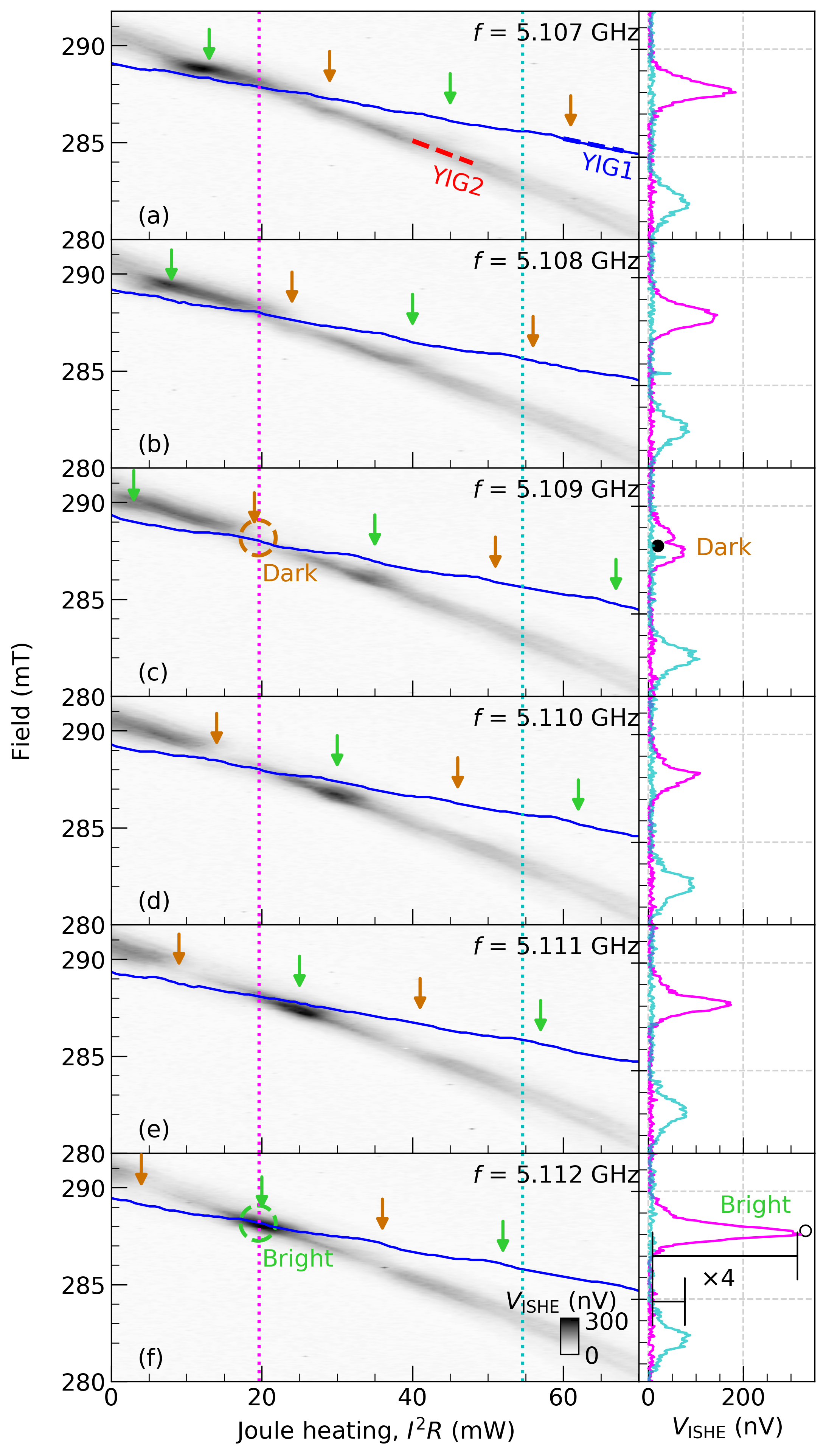}
  \caption{(Color online)  (a-f) Density plots of the spin pumping induced voltage, $V_\text{ISHE}$, in the Pt contact on the top YIG2 layer as a function of Joule heating and magnetic field observed at different frequency tuning relative to acoustic resonances (microwave power is set at 0~dBm and sample orientation is set at $\theta_{H}=15^{\circ}$). The blue solid lines are the maxima of the microwave absorption spectra that are dominated by YIG1. The right panel compares the modulation of the spin pumping signal measured either at zero detuning between the two Kittel resonances (purple vertical dotted line) or at large detuning (cyan vertical dotted line). The green and yellow arrows indicate the positions of odd and even phonon resonances, observable as small cusps in the microwave spectra (blue lines). The circles in (c) and (f) emphasize the triple resonance at which all frequencies agree.  (c) At the intersection with an even lattice mode {the peak splits} and becomes ``dark". (f) At the intersection with an odd mode, the intensity gains a factor of 4, which confirms that the magnetization amplitudes of both YIG1 and YIG2 interfere constructively into a ``bright" mode.}
  \label{fig_temp}
\end{figure}

We can then compute $P_2$ as a function of  the cooperativity, $\C=\Omega^2/(2\eta_s \eta_a)$, and $\rho \equiv h_1/h_2$, an asymmetry in the amplitudes of the external microwave magnetic fields at each layers, where $\ket{\varepsilon} = \mu_0 h_2 \,(\rho,0,1)/\sqrt{2}$.  At $\delta=0$, the contrast between bright and dark states follows the analytical expression:
\begin{equation}
\frac{P_{2,\text{bright}}}{ P_{2,\text{dark}}} =\left|\frac{(\rho+1) \C+2}{(\rho-1) \C-2}\right|^2. \label{eq:contrast}
\end{equation}
This ratio diverges for $\rho = 2/\C+1$, indicating that for any value of $\C$, there is an optimum coupling asymmetry $\rho$ to obtain the maximum contrast. In the asymptotic limit of $\C \gg 1$, we recover that the optimum is simply the symmetric system ($\rho\approx 1$).

\section{III. Experiments}

\begin{figure*}
  \includegraphics[width=1\textwidth]{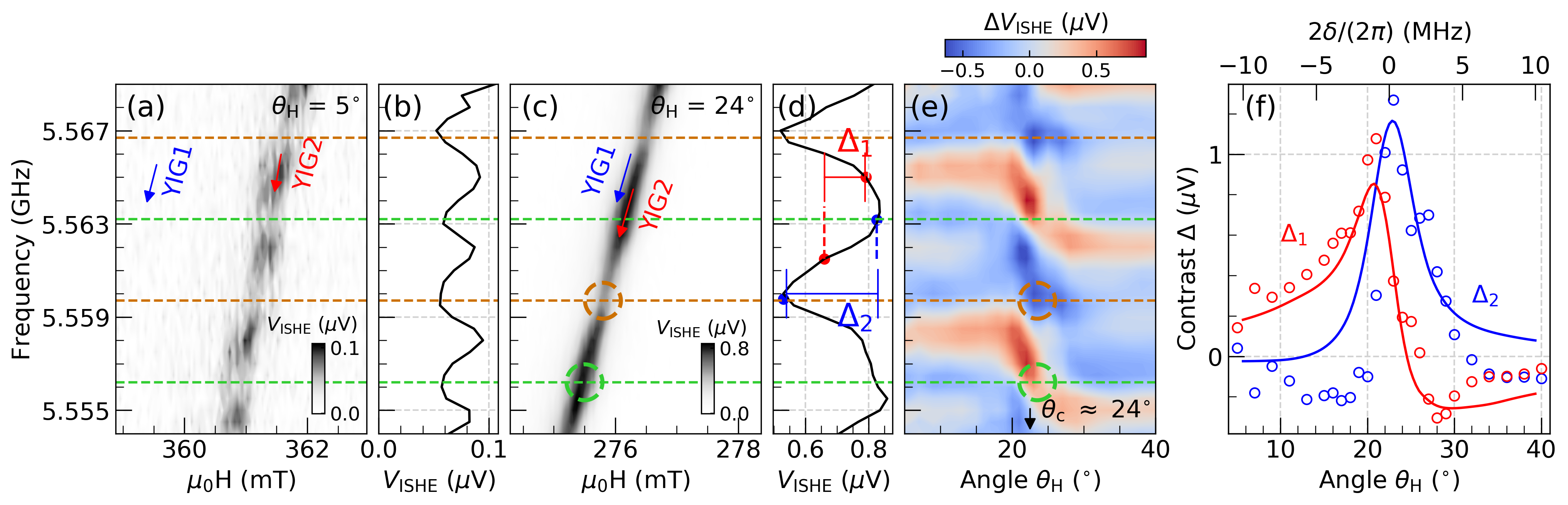}
  \caption{(Color online) (a,c) Gray-scale plot of the voltage signal $V_{\rm ISHE}$ generated by the Pt contact on YIG2 as a function of microwave frequency and magnetic field with tilt angles (a) $\theta_{H} = 5^{\circ}$ and (c) $\theta_{H}= 24^{\circ}$ (microwave power is set at 6 dBm). The blue arrow indicates the location of the Kittel mode of YIG1. (b,d) show the voltage modulation along the YIG2 resonance. The contrasts $\Delta_2$ (blue)  and $\Delta_1$ (red) in panel (d) are the voltage drops measured at the crossings between two consecutive phonon lines (blue dots) and at the middle of the crossings of two phonon lines (red dots), respectively, for $\theta_{H}= 24^{\circ}$. (e) Color-scale plot of $V_{\rm ISHE}$ as a function of frequency and polar angle. Here we subtract a constant background and correct for the angular dependence of the spin Hall effect (see text), where red/blue indicates positive/negative value. (f) The two different signals  $\Delta_1$ and $\Delta_2$ defined in panel (d) are plotted as a function of $\theta_{H}$. The solid lines are the expectations from the model.}
  \label{fig_angle}
\end{figure*}

We test and confirm the above model by room temperature spin pumping experiments with two nominally identical $d=200~$nm thick YIG films grown by liquid phase epitaxy on both sides of an $s=0.5$~mm thick GGG substrate as shown schematically in Fig.~\ref{fig_setup}(a). We denote the bottom and top YIG layers as YIG1 and YIG2, respectively. We monitor the YIG2 dynamics via the inverse spin Hall voltage ($V_\text{ISHE}$) \cite{saitoh2006conversion} across a 7~nm thick, 240~$\mu$m wide Pt, and 950~$\mu$m long Pt stripe  as shown in Fig.~\ref{fig_setup}(a). \footnote{The measured $V_\text{ISHE}$ signal vanishes at $\theta_H=0^\circ$, excluding a possible SSE contribution by in-plane thermal gradient induced by the inhomogeneous magnetic precession.} $V_\text{ISHE}$ peaks at $H_{\rm K2}$ as shown in Fig.~\ref{fig_setup}(b). We also measure the absorption power by a microwave antenna close to YIG1 as shown in Fig.~\ref{fig_setup}(c) for $\omega/(2\pi)=5.6$~GHz and  $\theta_H=+17^\circ$ \footnote{The cubic anisotropy breaks the inversion symmetry for rotation in the ($11\overline{2}$) plane relative to the [111] axis: $H_K^{\rm theory}(+\theta_H) \neq H_K^{\rm theory}(-\theta_H)$.}. 
At the resonance field $H_{\rm K1}$ the YIG1 Kittel mode dominates absorption by its proximity to the antenna, but we also see YIG2 by the weak line at $H_{\rm K2}$ that is shifted by 1.5~mT.  The asymmetric coupling ratio $\rho$ can be decreased by a larger separation between the microwave strip and the sample. The splitting between $H_{\rm K1}$ and $H_{\rm K2}$ betrays a small unintended asymmetry in the growth process that may be caused by a small difference in the saturation magnetization or magnetic anisotropies. The YIG2 linewidth is narrower in the $V_\text{ISHE}$ than in the microwave absorption spectra, which we attribute to a higher degree of homogeneity on the scale of the area covered by the Pt contact (see Appendix). We use below two different schemes  to accomplish the triple resonance condition with $H_{\rm K1}=H_{\rm K2}$.

Both the angle $\theta_{H}$ and a temperature gradient in the sample can tune the two Kittel modes. Fig.~\ref{fig_setup}(d) shows how the thin film demagnetization field shifts the resonance fields by more than 200 mT in the interval $\theta_{H}=0 \rightarrow \pm 90^\circ$ at 5.6 GHz. The result agree with Kittel formula, $H_{\rm K}^{\rm theory}$ \cite{Vonsovskii1966} (see Appendix) for rotations in the ($11\overline{2}$) plane, normal to the $[111]$ growth direction when parameterized by the uniaxial $H_{\rm U}$ and cubic $H_{\rm C}$ anisotropy fields. With $\mu_0 H_{\rm C2}=7.6$~mT, $\mu_0 H_{\rm U2} = 3.5$~mT, the gyromagnetic ratio of $\gamma_2/(2\pi) = 28.5$~MHz/mT and magnetization $\mu_{\rm 0}M_2=0.172$~T, we fit the top layer $H_{\rm K2}(\theta_{H})$ measurements  (solid line in Fig.~\ref{fig_setup}(d)) . The small deviations of  less than $\pm 0.3$\% plotted in Fig.\ref{fig_setup}(e) show a crossover angle $\theta_{\rm c} \approx 24^{\circ}$ as indicated by an arrow. 

The magnetic resonance conditions depend on temperature, so a temperature gradient over the stack can tune $H_{\rm K1}-H_{\rm K2}$, with the advantage that each of $H_{\rm K}$'s varies much less than for the angular scan. We can heat the system simply by the Ohmic dissipation of a dc current $I$ in the Pt contact. Since the dc source increases the noise level, we tune to the cavity resonance frequency, which has a quality factor of about 30. The higher power enhances the signal to noise ratio of the spin pumping signal. We detect the spin pumping voltage by lock-in at a frequency of about 3.1 kHz, thereby removing parasitic dc signals. The spin Seebeck voltage induced by microwave heating is negligibly small (see Appendix). At the magnetic field tilt angle $\theta_H = 15^{\circ}$ the two resonances are so close that they cross already at small temperature gradients. 
In Fig.~\ref{fig_setup}(f), the resonance fields from both the top and bottom layers are plotted as a function of the Joule heating power $I^2 R$. 
Since the magnetization decreases linearly with temperature, the accompanied reduction of the resonance field can be used as a thermometer \cite{an2020short}. From the shift of the YIG2 and YIG1 resonance fields, we estimate a temperature increase of 61 K and 31 K for $I = 26 \, \mathrm{mA}$ ($I^2R\sim70$~mW), respectively \footnote{From the 11\% relative increase of the Pt resistance, we infer a temperature rise of 52~K at the top layer at the same current, 26~mA, 
giving a second estimate in agreement within 15\%.}.  The crossover current $I_{\rm c}$ is at $\sim$13.6 mA, corresponding to Joule heating of $I_{\rm c}^2R\approx  19~\rm mW$. Higher temperature expands the crystal and reduces the sound velocity, so the phonon frequencies shift as a function of $T$.

\subsection{Tuning by temperature gradients}

\begin{figure*}
  \includegraphics[width=1\textwidth]{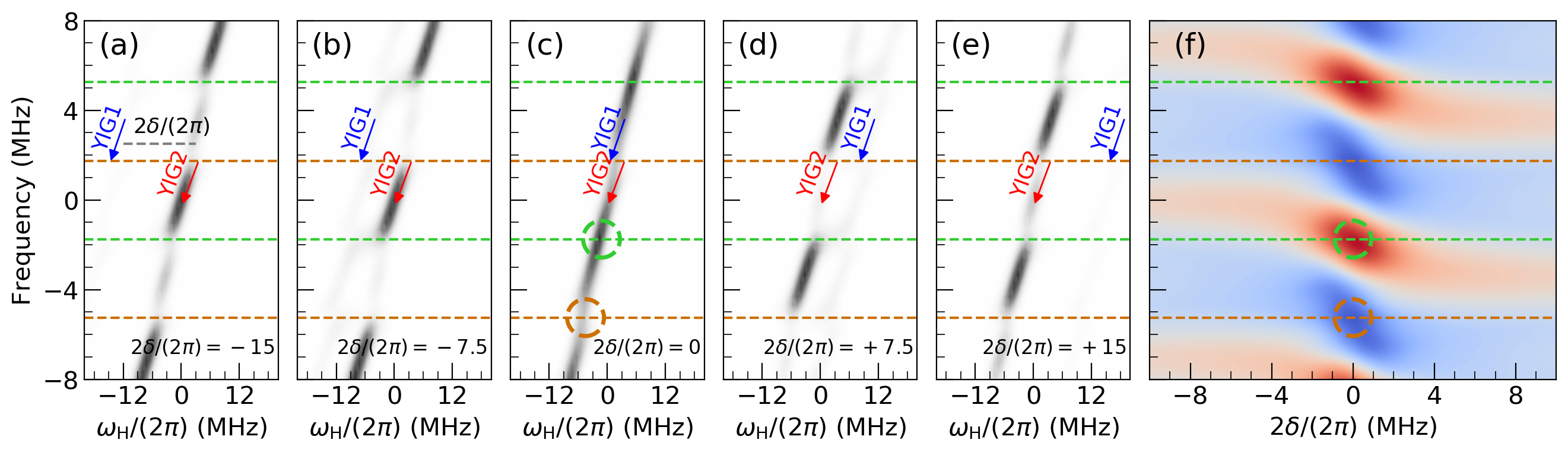}
  \caption{(Color online) (a-e) Numerically calculated intensity $P_2(\omega)$ plotted for different detunings $\delta$ between top and bottom Kittel resonance frequencies (parameters given in the text). The abscissa $\omega_H = \gamma H$ corresponds to a field sweep inducing a shift of the two Kittel modes along the resonance condition, while the even (orange dash) or odd (green dash) phonons do not depend on the field (horizontal lines). The gray scale indicates the calculated power dissipated by $m_2$. (f) Calculated dependence of the maximum signal as a function of the frequency and the frequency difference, $2\delta / (2 \pi)$.}
  \label{fig_cal}
\end{figure*}

In Fig.~\ref{fig_temp}(a-f), we plot $V_{\rm ISHE}$ as a function of $I^2R$ and magnetic field at six different frequencies $f$ in the grey scale indicated in the bottom panel. We average the data for positive and negative currents to remove effects of the current direction  such as spin-orbit torques or bolometric voltages \cite{hahn2013comparative}. We overlay $V_{\rm ISHE}$  by blue solid lines that are the maxima of the microwave absorption spectra as a function of magnetic field, i.e. the resonances of the bottom layer. Small kinks emphasized by the green and orange arrows are evidence for crossings of the magnon mode with odd and even phonon modes, respectively \footnote{We assign the phonon resonances to the dips in the integrated intensity of the microwave absorption spectrum (see Appendix)}. The decrease of the ultrasound velocities  $v$ \cite{an2016magnons} causes a downward shift of the phonon frequencies with increasing temperature \footnote{The phonon frequency shift is determined by the average temperature of the whole stack, but a small temperature gradient has no effect to leading order in the perturbation.}. The distortions are expected for $f = i\, v(1-\zeta I^2R)/[2(2d+s)]$, where $i$ is the number of nodes in the standing wave acoustic phonon amplitude ($i$ is of the order of 1400 here). We indeed observe abnormalities for different indices $i$ and heating powers $I^2R$, so several phonon modes contribute to the window opened by the (moderate) temperature gradient generated by the $\leq$ 70 mW heating power. The decrease in $v$ is proportional to the heating power with coefficient  $\zeta=0.048~\rm mW^{-1}$ as estimate from the five phonon lines that shift by $3.5 \times 5 = 17.5$~MHz.

While the peak position of $V_{\rm ISHE}$ decreases linearly with temperature, similar to the microwave signal of the bottom layer, its amplitude changes dramatically when  $H_{\rm K1} \approx$ $H_{\rm K2}$. The signals vary strongly in the power range of 0-40 mW, but saturate in the 40 - 70 mW interval. In Fig.~\ref{fig_temp}(f) we observe an enhanced peak by more than a factor of 2 when an odd phonon mode comes close to the crossing at $I=I_{\rm c}$ indicated by the vertical purple-dotted line. On the other hand, according to Fig.~\ref{fig_temp}(c) an even phonon mode at $I=I_{\rm c}$ splits the peak. In our experiments, $m_1$ is excited stronger than $m_2$ by its proximity to the antenna, while a complete dark state emerges only when the amplitudes of $m_1$ and $m_2$ are the same, as shown in Fig.~\ref{fig_model}. At a current far from the crossover $I_c$, indicated by the vertical dotted cyan line and the corresponding line cut in the right panels of Fig.~\ref{fig_temp}, such features are absent. The $V_{\rm ISHE}$ signal at the peak center of the purple curves is enhanced by a factor of 4 (attenuated by a factor of 2) when an odd (even) acoustic mode mediates the coupling.  

These factors can be compared to the predictions by Eq.(\ref{eq:contrast}). While the value of $\C \approx 6$ (see below), which depends on the material parameters, is fixed, the value of $\rho$ is adjusted depending on the sample mounting (spacer between the garnet sample and the microwave antenna) and the power used. According to Eq.(\ref{eq:contrast}), such ratio between bright and dark state corresponds to an asymmetric inductive coupling $\rho$ between YIG1 and YIG2 close to 3, which is smaller than previous estimation \cite{an2020coherent}.  
We attribute this partially to the use of large power, which saturates the excitation of the bottom layer while the top layer is still in the linear regime at the chosen microwave input power.

\subsection{Tuning by magnetic field angle}

Next we interpret the $V_\text{ISHE}$ spectra as a function of the polar angle $\theta_H$ without applying any current to the Pt. Figure~\ref{fig_angle}(a) is a gray-scale density plot of $V_\text{ISHE}$ versus frequency and magnetic field close to the normal configuration ($\theta_H=5^\circ$), with large differences between the resonance frequency of the two layers, indicated by the  of blue and red arrows. Green and orange horizontal lines mark the odd and even phonon resonances, respectively, that are spaced by 3.5 MHz. At large detuning, as shown in Fig.~\ref{fig_angle}(a), magnon-phonon hybridization just reduces the spin-pumping voltage. This can be interpreted as  an increased broadening that indicates an enhanced magnetization damping by leakage into the lattice \cite{an2020coherent}. 

Around $\theta_H=24^\circ$ the magnetic resonance of both layers are very close, as indicated by the blue and red arrows in Fig.~\ref{fig_angle}(c), which looks very different from panel (a). The minima at the green lines for $\theta_H=5^\circ$ evolve into maxima, while the minima at the orange lines become deeper. We highlight these features in the adjacent panels (b) and (d) in terms of the signals plotted along the resonance condition. $\Delta_2$ in panel (d) is the difference between the signals at two adjacent odd and even phonon crossings (blue points), while the difference between the voltages shifted by a quarter of a phonon wavelength from the crossing (red points) is $\Delta_1$.  

Figure \ref{fig_angle}(e) is a color density plot of $\Delta V_{\rm ISHE} \equiv (V_{\rm ISHE}-\overline{V}_{\rm ISHE})/\sin{\theta_{M}}$ as a function of $\theta_H$, where $\theta_M$ is the calculated angle between the magnetization and film normal. The factor $\sin \theta_M$  eliminates the well-known angle dependence of the inverse spin Hall effect  \cite{ando2008angular}. Furthermore, by subtracting  $\overline{V}_{\rm ISHE}$, i.e. the average over the frequency range at a fixed angle, we remove possible deviations from the $\sin \theta_M$ dependence caused by an elliptical precession \cite{ando2011inverse}.
Figure~\ref{fig_angle}(f) emphasizes that the modulation either changes sign near the crossover angle $\theta_{\rm c}$ or develops a maximum, which agrees with the expectations from the 3-oscillator model. These shapes may be interpreted in terms of the real and imaginary parts of a dynamic susceptibility of the coupled magnetic layers.

Figure~\ref{fig_cal} summarizes in more detail the numerical solutions of the 3 coupled equations for $\Omega/(2\pi)$  = 1.5 MHz, $\eta_s/(2\pi)$ = 0.5 MHz, $\eta_a/(2\pi)$ = 0.35 MHz, which lead to $\C \approx 6$. We use $\rho \approx 7$, a value close to our low power estimation \cite{an2020coherent}. Even though our  model is strictly linear, it captures the qualitative features well even at the high applied microwave powers \footnote{The microwave power for the temperature-tuning experiments  was about 7 times higher than that use in the angle-tuning measurement. The clear nonlinear features such as asymmetric line shapes in the former experiments are caused by the individual magnetic layer. The coupling is still a small perturbation, so the observed features should follow a linear model.}. Both Figs.~\ref{fig_angle}(e,f) and Fig.~\ref{fig_cal}(f)) find a double sign change of the signal when crossing the triple resonance. When the detuning between the two Kittel modes vanishes in Fig.~\ref{fig_cal}(c), the bright and dark states appear at the triple crossing indicated by green and dark orange circles. In theory, the bright and dark states appear exactly at the phonon frequencies. The measured extrema in Fig.~\ref{fig_angle}(e) slightly deviate from the predictions  because at the degeneracy points the signals are sensitive to higher order corrections. In the fully hybridized regime, the pure bright and dark states emerge as shown in Fig.~\ref{fig_model}(g) and (h) only when $\rho = 1$, i.e. when the two layers are evenly excited by the microwaves. If $\rho \ne 1$, then the contrast in the dynamics of $m_2$ is reduced, leading to a deviation from the model presented in Fig.\ref{fig_model}. We do not observe any evidence for damping-mediated coupling that induces level attraction rather than repulsion \cite{heinrich2003dynamic,harder2018level,li2020coherent}. This is an interesting subject for future studies.

\section{IV. Conclusions}

We demonstrate coherent coupling of two macrospins by phonon exchange over a millimeter distance in YIG\textbar GGG \textbar YIG phononic spin valves. The phenomenology is generic for coupled tri-partite systems that support bonding and anti-bonding states at the degeneracy points. 
The additional knob to switch the polarity of the mutual coupling through relatively small changes of the mechanical resonator (\textit{e.g.} $\sim 0.1\%$ thickness variations) drastically tunes the inductive coupling of a magnetic system by ultrasound, and allows an integrated solution for long-range communication between distant quantum states. Modulation at short time scales (shorter than the relaxation times but longer than the reciprocal coupling rates) should allow coherent permutation between bright and dark states by adiabatic fast passage \cite{carroll1986transition,militello2019three,schwienbacher2020dia}. This effect could protect quantum information by switching on and off the radiation damping to an external readout antenna. While our paper focuses essentially on the dark or bright states, where the coupled dynamics of the two macrospins are respectively quenched or enhanced  by the coherent sum of the two contributions, a continuous control of the mutual phase between the two oscillators measured by a phase-resolved technique
could be beneficial to better quantify the coherent coupling phenomena.

\section{Appendices}
\subsection{Appendix A: Spin Seebeck voltage driven by microwave heating} 
\begin{figure}[htp]
    \includegraphics[width=0.5\textwidth]{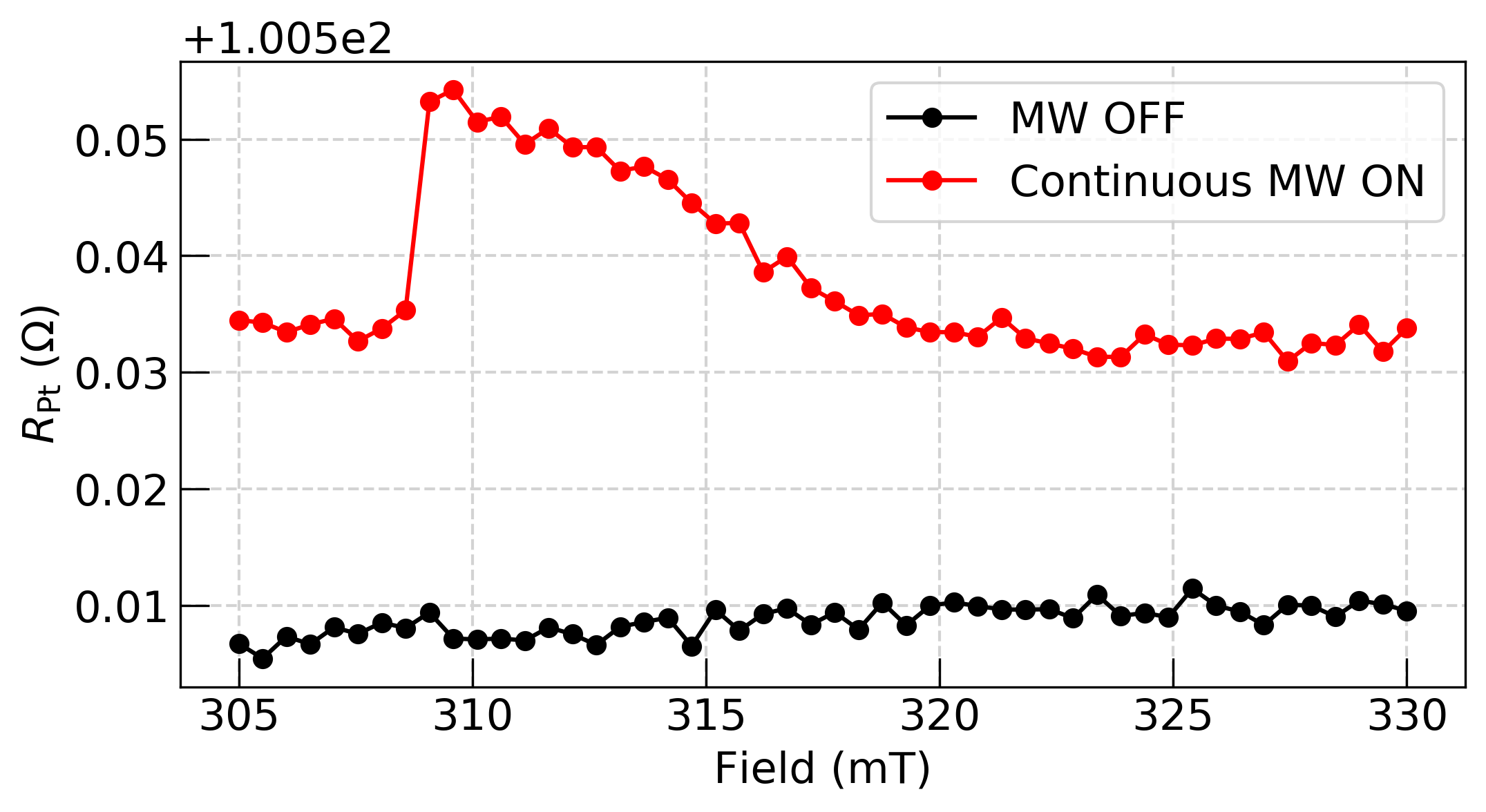}
    \centering
    \caption{Measured $R_{\rm Pt}$ under continuous microwave excitation ON (red) and OFF (black) at 6 dBm and 5.11 GHz. The background level increases by 25 $\rm m\Omega$ due to the microwave heating. In addition to that, the magnetic resonance, represented by the foldover shape on the red curve due to the strong excitation power, also increases $R_{\rm Pt}$.}
    \label{fig_heat}
\end{figure}

The microwave heating can give rise to a spin Seebeck effect which induces spin currents and therefore an additional voltage in Pt may arise in the lock-in signal. We estimate a temperature rise of about 100 mK for a continuous microwave heating based on the increase of resistance as shown in Fig.~\ref{fig_heat}, with $\Delta R=R_0 \zeta \Delta T$, where $\zeta = 0.0021~\rm K^{-1}$ for Pt. Using the standard SSE coefficient of $S=0.08 \rm~\mu V/K$ for YIG$|$Pt system \cite{guo2016influence}, the Pt length of $L_{\rm Pt}=0.95~\rm mm$, and the SSE voltage expression of $V_{\rm SSE} = S L_{\rm Pt} \Delta T/(s+2d)$, we obtain the SSE voltage of about 15 nV, which is close to the uncertainty of the measured spin pumping voltage and may be safely disregarded.

\subsection{Appendix B: Angular dependence of the Kittel formula} \label{App_Kittel}
We fit the angular dependence of the resonance field by the following formula that includes the uniaxial $H_{\rm U}$ and cubic $H_{\rm C}$ anisotropy fields \cite{Vonsovskii1966}.

\begin{multline}
\left(\frac{\omega_K}{\gamma\mu_0}\right)^2\sin^2{\theta_M}=\Big[
H_K \cos{(\theta_H-\theta_M)}+ (H_{\rm U}-M)\cos{2\theta_M}\\
+ H_{\rm C}\Big(-\frac{7}{12}\cos{4\theta_M}-\frac{1}{12}\cos{2\theta_M}\\
+\frac{\sqrt{2}}{6}\sin{2\theta_M}(3-8\sin^2{\theta_M})\Big)\Big] \\
\times\Big[H_K\sin{\theta_H}\sin{\theta_M}-\frac{3}{\sqrt{2}}H_{\rm C}\sin^3{\theta_M}\cos{\theta_M}\Big],
\end{multline}

where $\theta_H$ and $\theta_M$ represent the directions of the magnetic field and magnetization, respectively. $\theta_H$ and $\theta_M$ are related according to:

\begin{multline}
0=H_K \sin{(\theta_M-\theta_H)}+\left(\frac{H_{\rm U}-M}{2}\right)\sin{2\theta_M}\\
-\frac{1}{24}H_{\rm C}\big[\sin{2\theta_M}(1+7\cos{2\theta_M})-2\sqrt{2}(\cos{2\theta_M}-\cos{4\theta_M})\big].
\end{multline}

Our results agree well with the above formula. The fit for the top layer $H_{\rm K2}(\theta_{H})$ leads to the  parameters $\mu_0 H_{\rm C2}=7.6$~mT, $\mu_0 H_{\rm U2} = 3.5$~mT, the gyromagnetic ratio of $\gamma_2/(2\pi) = 28.5$~MHz/mT, and magnetization $\mu_{\rm 0}M_2=0.172$~T (solid line in Fig.~\ref{fig_setup}(d)). 

\subsection{Appendix C: Identification of phonon resonances} \label{App_Iden}

\begin{figure}[htp]
    \includegraphics[width=0.5\textwidth]{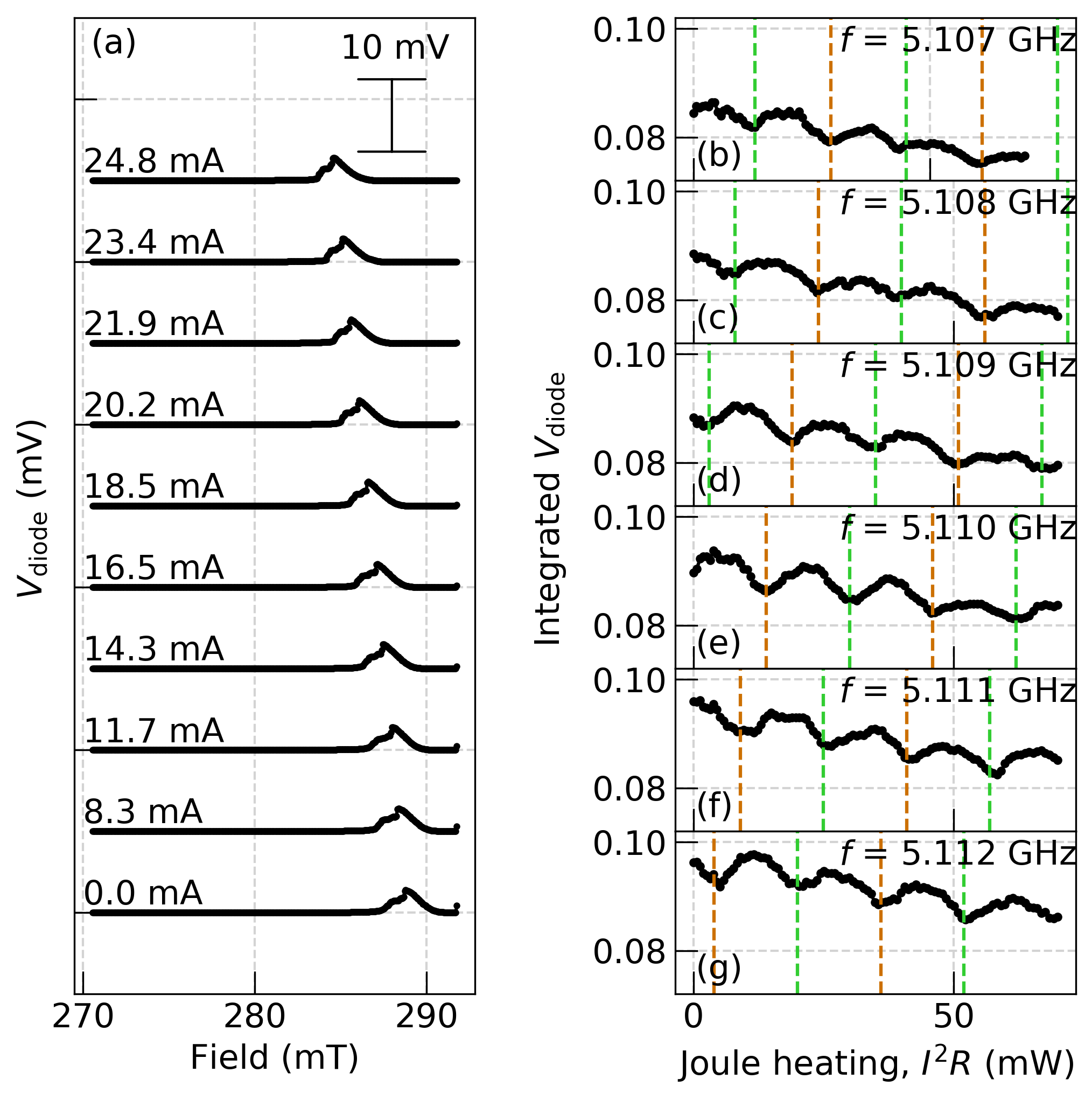}
    \centering
    \caption{(a) The diode absorption voltage $V_{\rm diode}$ measured simultaneously with $V_{\mathrm{ISHE}}$ during the temperature-tuning experiment. Data are shown for $f=5.11~\rm GHz$ with different current values. (b-g) Integrated $V_{\rm diode}$ as a function of Joule heating for different frequencies. The green and dark orange lines represent the odd and even phonons.}
    \label{fig_diode}
\end{figure}

The frequencies of the standing phonon states decrease with increasing temperature due to Joule heating caused by increasing current. We trace their positions systematically by the integrated intensity of the microwave absorption signal. Fig.~\ref{fig_diode}(a) shows the shift of the raw absorption spectra. Periodic dips in the integrated signals as a function of Joule heating reveal the phonon resonances in Fig.~\ref{fig_diode}(b-g).

\subsection{Appendix D: Detailed comparison of magnetic resonance spectra obtained from spin pumping and diode absorption methods} \label{App_Mag}

\begin{figure}[hth]
\includegraphics[width=0.5\textwidth]{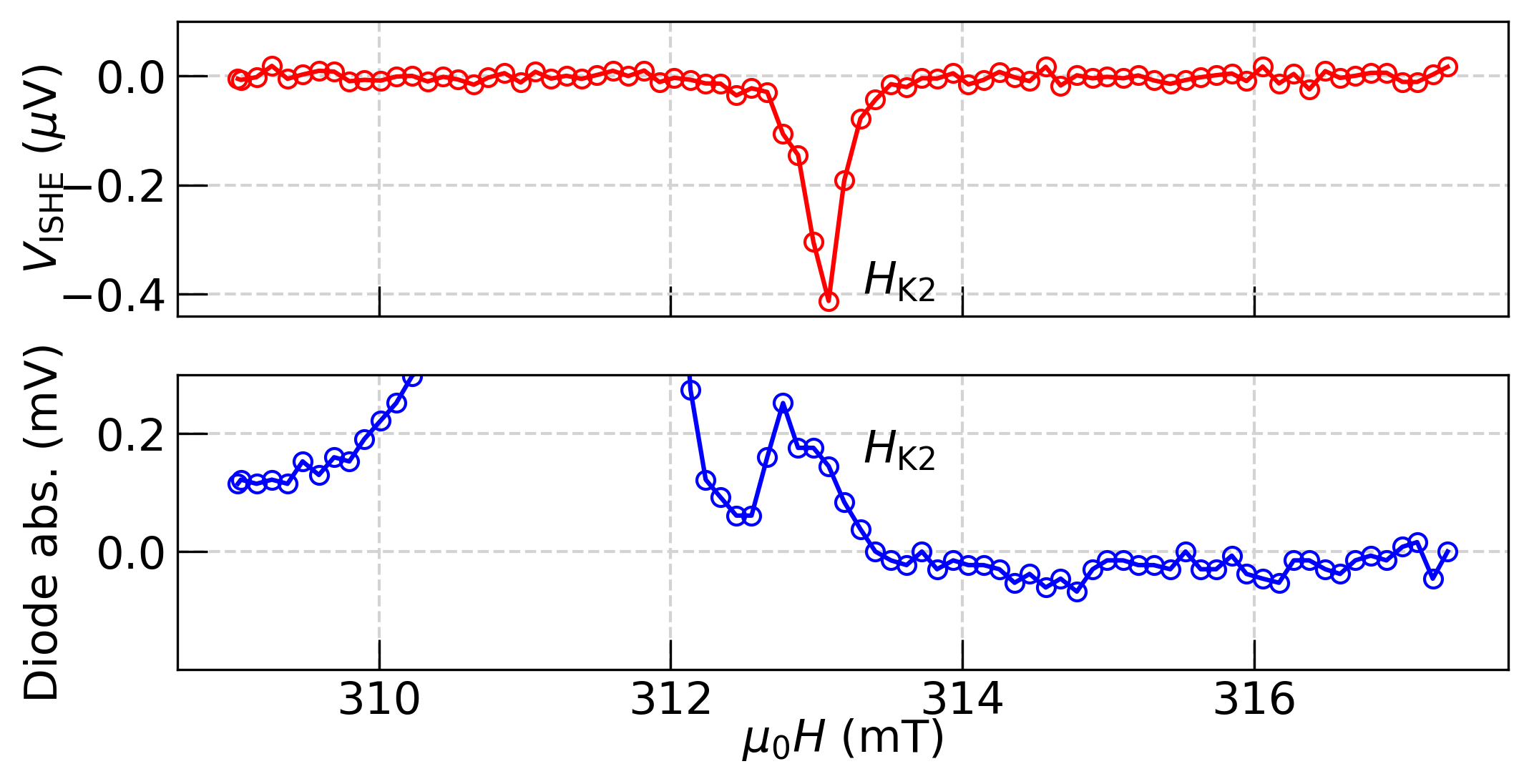}
\centering
\caption{Comparison of $H_{\mathrm{K2}}$ measured from $V_{\mathrm{ISHE}}$
(top) and diode (bottom). }%
\label{figR_HK2}%
\end{figure}

For the sake of completeness, we provide in Fig.~\ref{figR_HK2} a zoom around of the Kittel mode of the YIG2 layer of both the spin pumping and diode absorption spectra as shown in Fig.\ref{fig_setup} (b) and (c). It shows that indeed the YIG2 resonances positions are slightly shifted (by 0.3 mT) between the two methods. Also the linewidth measured from the spin pumping method is narrower than that from the diode absorption. We attribute this to the inhomogeneous broadening, where in one case the spin pumping is sensitive to the spins below the Pt surface, while in the other case the inductive signal is sensitive to the bulk magnetization dynamics.

\section*{Acknowledgements}
This work was partially supported by the French Grants ANR-18-CE24-0021 Maestro and ANR-21-CE24-0031 Harmony; the EU-project HORIZON-EIC-2021-PATHFINDEROPEN-01 PALANTIRI-101046630. K.A. acknowledges support from the National Research Foundation of Korea (NRF) grant (No. 2021R1C1C201226911) funded by the Korean government (MSIT). The JSPS supported G.B. by Kakenhi Grant No. 19H00645. Additionnaly, this work was partially supported  by the U.S. NSF under the grant No. EFMA-1641989, by the U.S. Air Force Office of Scientific Research under the MURI Grant No. FA9550-19-1-0307, by the DARPA TWEED Program under the grant DARPA-PA-19-04-05-FP-001 and by the Oakland University Foundation.

%\bibliography{bib}

\begin{thebibliography}{68}
\bibitem [{fn2()}]{fn2}%
  \BibitemOpen
  \href@noop {} {}\bibinfo {note} {The $\sim 0.5$~m{T} difference of resonance
  fields between $\pm 90^\circ$ observed in the inset of Fig.2(d) is attributed
  to a lateral shift of the sample position with respect to the center of the
  magnet when rotating by 180$^\circ$}\BibitemShut {NoStop}%
\end{thebibliography}


\end{document}